\documentclass[aps,prd,altaffilletter,reprint,amsmath,amssymb,
longbibliography,
superscriptaddress,showpacs]{revtex4-1}
\usepackage{graphicx}
\usepackage{dcolumn}
\usepackage{bm}
\usepackage{bbding}
\usepackage{multirow}
\usepackage{color}
\usepackage{lineno}

\begin{document}

\title{Search for the rare decay $K^+\to\mu^+\nu\bar\nu\nu$}

\author{A.V.~Artamonov}\affiliation{Institute for High Energy Physics,
Protvino,
Moscow Region, 142 280, Russia}
\author{B.~Bassalleck}\affiliation{Department of Physics and Astronomy,
University of New Mexico, Albuquerque, NM 87131}
\author{B.~Bhuyan}
\altaffiliation{Now at Department of Physics, Indian Institute of
Technology Guwahati, Guwahati, Assam, 781 039, India.}
\affiliation{Brookhaven
National Laboratory, Upton, NY 11973}
\author{E.W.~Blackmore}\affiliation{TRIUMF, 4004 Wesbrook Mall, Vancouver,
British Columbia, Canada V6T 2A3}
\author{D.A.~Bryman} \affiliation{Department of Physics and Astronomy,
University of British Columbia, Vancouver, British Columbia, Canada V6T 1Z1}
\author{S.~Chen} \affiliation{Department of Engineering Physics, Tsinghua
University, Beijing 100084, China} \affiliation{TRIUMF, 4004 Wesbrook Mall,
Vancouver, British Columbia, Canada V6T 2A3}
\author{I-H.~Chiang} \affiliation{Brookhaven National Laboratory, Upton, NY
11973}
\author{I.-A.~Christidi}
\altaffiliation{Now at Physics Department, Aristotle
University of Thessaloniki, Thessaloniki 54124, Greece.}
\affiliation{Department of
Physics and Astronomy, Stony Brook University, Stony Brook, NY 11794}
\author{P.S.~Cooper}\affiliation{Fermi National Accelerator Laboratory,
Batavia, IL 60510}
\author{M.V.~Diwan} \affiliation{Brookhaven National Laboratory, Upton, NY
11973}
\author{J.S.~Frank}
\altaffiliation{Now at 1 Nathan Hale Drive, Setauket, New York 11733.}
\affiliation{Brookhaven National Laboratory, Upton, NY 11973}
\author{T.~Fujiwara}\affiliation{Department of Physics, Kyoto University,
Sakyo-ku, Kyoto 606-8502, Japan}
\author{J.~Hu} \affiliation{TRIUMF, 4004 Wesbrook Mall, Vancouver, British
Columbia, Canada V6T 2A3}
\author{J.~Ives} \affiliation{Department of Physics and Astronomy, University
of British Columbia, Vancouver, British Columbia, Canada V6T 1Z1}
\author{A.O.~Izmaylov}\affiliation{Institute for Nuclear Research RAS, 60
October Revolution Prospect 7a, 117312 Moscow, Russia}
\author{D.E.~Jaffe} \affiliation{Brookhaven National Laboratory, Upton, NY
11973}
\author{S.~Kabe}
\altaffiliation{Deceased.}
\affiliation{High Energy Accelerator Research Organization~(KEK), Oho, Tsukuba,
Ibaraki 305-0801, Japan}
\author{S.H.~Kettell} \affiliation{Brookhaven National Laboratory, Upton, NY
11973}
\author{M.M.~Khabibullin}\affiliation{Institute for Nuclear Research RAS, 60
October Revolution Prospect 7a, 117312 Moscow, Russia}
\author{A.N.~Khotjantsev}\affiliation{Institute for Nuclear Research RAS, 60
October Revolution Prospect 7a, 117312 Moscow, Russia}
\author{P.~Kitching} \affiliation{Centre for Subatomic Research, University of
Alberta, Edmonton, Canada T6G 2N5}
\author{M.~Kobayashi}
\affiliation{High Energy Accelerator Research Organization~(KEK), Oho, Tsukuba,
Ibaraki 305-0801, Japan}
\author{T.K.~Komatsubara}
\affiliation{High Energy Accelerator Research Organization~(KEK), Oho, Tsukuba,
Ibaraki 305-0801, Japan}
\author{A.~Konaka} \affiliation{TRIUMF, 4004 Wesbrook Mall, Vancouver, British
Columbia, Canada V6T 2A3}
\author{Yu.G.~Kudenko}\affiliation{Institute for Nuclear Research RAS, 60
October Revolution Prospect 7a, 117312 Moscow, Russia}\affiliation{Moscow
Institute of
Physics and Technology, 141700 Moscow, Russia}
\affiliation{National Research Nuclear University MEPhI (Moscow Engineering
Physics Institute), 115409 Moscow, Russia}
\author{L.G.~Landsberg}\altaffiliation{Deceased.}\affiliation{Institute for
High Energy Physics, Protvino, Moscow Region, 142 280, Russia}
\author{B.~Lewis}\affiliation{Department of Physics and Astronomy, University
of New Mexico, Albuquerque, NM 87131}
\author{K.K.~Li}\affiliation{Brookhaven National Laboratory, Upton, NY 11973}
\author{L.S.~Littenberg} \affiliation{Brookhaven National Laboratory, Upton, NY
11973}
\author{J.A.~Macdonald} \altaffiliation{Deceased.} \affiliation{TRIUMF, 4004
Wesbrook Mall, Vancouver, British Columbia, Canada V6T 2A3}
\author{J.~Mildenberger} \affiliation{TRIUMF, 4004 Wesbrook Mall, Vancouver,
British Columbia, Canada V6T 2A3}
\author{O.V.~Mineev}\affiliation{Institute for Nuclear Research RAS, 60 October
Revolution Prospect 7a, 117312 Moscow, Russia}
\author{M. Miyajima} \affiliation{Department of Applied Physics, Fukui
University, 3-9-1 Bunkyo, Fukui, Fukui 910-8507, Japan}
\author{K.~Mizouchi}\affiliation{Department of Physics, Kyoto University,
Sakyo-ku, Kyoto 606-8502, Japan}
\author{N.~Muramatsu}\altaffiliation{Now at Research Center for Electron Photon
Science, Tohoku University, Taihaku-ku, Sendai, Miyagi 982-0826,
Japan.}\affiliation{Research Center for Nuclear Physics, Osaka University, 10-1
Mihogaoka, Ibaraki, Osaka 567-0047, Japan}
\author{T.~Nakano}\affiliation{Research Center for Nuclear Physics, Osaka
University, 10-1 Mihogaoka, Ibaraki, Osaka 567-0047, Japan}
\author{M.~Nomachi}\affiliation{Laboratory of Nuclear Studies, Osaka
University, 1-1 Machikaneyama, Toyonaka, Osaka 560-0043, Japan}
\author{T.~Nomura}\altaffiliation{Now at High Energy Accelerator Research
Organization (KEK), Oho, Tsukuba, Ibaraki 305-0801,
Japan.}\affiliation{Department of Physics, Kyoto University, Sakyo-ku,
Kyoto 606-8502, Japan}
\author{T.~Numao} \affiliation{TRIUMF, 4004 Wesbrook Mall, Vancouver, British
Columbia, Canada V6T 2A3}
\author{V.F.~Obraztsov}\affiliation{Institute for High Energy Physics,
Protvino, Moscow Region, 142 280, Russia}
\author{K.~Omata}
\affiliation{High Energy Accelerator Research Organization~(KEK), Oho, Tsukuba,
Ibaraki 305-0801, Japan}
\author{D.I.~Patalakha}\affiliation{Institute for High Energy Physics,
Protvino, Moscow Region, 142 280, Russia}
\author{R.~Poutissou} \affiliation{TRIUMF, 4004 Wesbrook Mall, Vancouver,
British Columbia, Canada V6T 2A3}
\author{G.~Redlinger} \affiliation{Brookhaven National Laboratory, Upton, NY
11973}
\author{T.~Sato}
\affiliation{High Energy Accelerator Research Organization~(KEK), Oho, Tsukuba,
Ibaraki 305-0801, Japan}
\author{T.~Sekiguchi}
\affiliation{High Energy Accelerator Research Organization~(KEK), Oho, Tsukuba,
Ibaraki 305-0801, Japan}
\author{A.T.~Shaikhiev}\affiliation{Institute for Nuclear Research RAS, 60
October Revolution Prospect 7a, 117312 Moscow, Russia}
\author{T.~Shinkawa} \affiliation{Department of Applied Physics, National
Defense Academy, Yokosuka, Kanagawa 239-8686, Japan}
\author{R.C.~Strand} \affiliation{Brookhaven National Laboratory, Upton, NY
11973}
\author{S.~Sugimoto} \altaffiliation{Deceased.}
\affiliation{High Energy Accelerator Research Organization~(KEK), Oho, Tsukuba,
Ibaraki 305-0801, Japan}
\author{Y.~Tamagawa} \affiliation{Department of Applied Physics, Fukui
University, 3-9-1 Bunkyo, Fukui, Fukui 910-8507, Japan}
\author{R.~Tschirhart} \affiliation{Fermi National Accelerator Laboratory,
Batavia, IL 60510}
\author{T.~Tsunemi} \altaffiliation{Now at Department of Physics, Kyoto
University, Sakyo-ku, Kyoto 606-8502, Japan.}
\affiliation{High Energy Accelerator Research Organization~(KEK), Oho, Tsukuba,
Ibaraki 305-0801, Japan}
\author{D.V.~Vavilov} \altaffiliation{Now at TRIUMF, 4004 Wesbrook Mall,
Vancouver, British Columbia, Canada V6T 2A3.}\affiliation{Institute for High
Energy Physics, Protvino, Moscow Region, 142 280, Russia}
\author{B.~Viren} \affiliation{Brookhaven National Laboratory, Upton, NY 11973}
\author{Zhe~Wang} \affiliation{Department of Engineering Physics, Tsinghua
University, Beijing 100084, China} \affiliation{Brookhaven National Laboratory,
Upton, NY 11973}
\author{Hanyu~Wei} \affiliation{Department of Engineering Physics, Tsinghua
University, Beijing 100084, China}
\author{N.V.~Yershov}\affiliation{Institute for Nuclear Research RAS, 60
October Revolution Prospect 7a, 117312 Moscow, Russia}
\author{Y.~Yoshimura}
\affiliation{High Energy Accelerator Research Organization~(KEK), Oho, Tsukuba,
Ibaraki 305-0801, Japan}
\author{T.~Yoshioka}
\altaffiliation{Now at  Department of Physics, Kyushu University, Higashi-ku,
Fukuoka 812-8581, Japan.}
\affiliation{High Energy Accelerator Research Organization~(KEK), Oho, Tsukuba,
Ibaraki 305-0801, Japan}
\collaboration{E949 Collaboration}\noaffiliation

\date{\today}

\begin{abstract}

Evidence of the $K^+\to\mu^+\nu\bar\nu\nu$ decay was searched for using E949~(Brookhaven 
National Laboratory, USA) experimental  data 
with an exposure of $1.70\times 10^{12}$ stopped kaons. The data sample is dominated by the 
background process $K^+\to\mu^+\nu_\mu\gamma$. 
An upper limit on the decay rate $\Gamma(K^+\to\mu^+\nu\bar\nu\nu)< 2.4\times
10^{-6}\Gamma(K^+\to \textnormal{all})$ at 90\% confidence level was set assuming the Standard 
Model muon
spectrum. The data are presented in
such a way as to allow calculation of rates for any assumed $\mu^+$ spectrum.   

\end{abstract}


\maketitle

\section{Introduction \label{sec:intro}}
The $K^+\to\mu^+\nu\bar\nu\nu$ decay involves four fermions and cannot occur in first order in the 
Standard Model~(SM). Therefore, its investigation provides information about higher-order weak 
effects.  
The most recent calculation of this process in the framework of the SM has been done by 
D.~Gorbunov and A.~Mitrofanov~\cite{StandardModel}. The SM predicts an extremely low total rate 
for this process~($\mathcal{O}(10^{-16})$).

\begin{figure}[]
\centering
\includegraphics[width=\columnwidth]{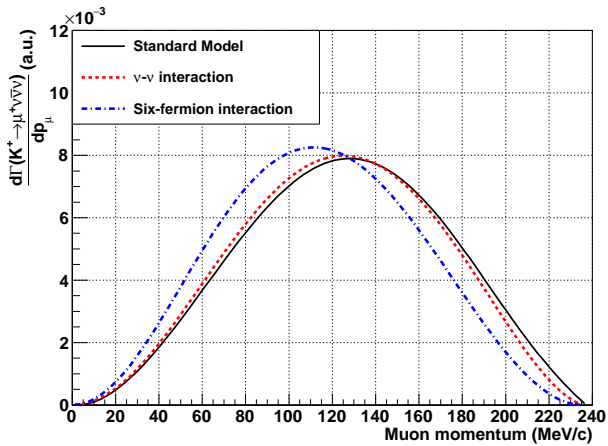}
\caption{Differential decay rates for $K^+\to\mu^+\nu\bar\nu\nu$ process in the framework of the 
Standard Model, $\nu-\nu$ interaction model with $F=1$, and six fermion interaction model with 
$F_S=1$. All spectra are normalized to 1.}
\label{fig:models}
\end{figure}

The study of the $K^+\to\mu^+\nu\bar\nu\nu$ decay can also provide information on two effects: a 
neutrino-neutrino interaction~\cite{nunu,nununew} and  a six-fermion interaction~\cite{sixf,sixf2}. The 
differential decay rates for $K^+\to\mu^+\nu\bar\nu\nu$ are the following:
\begin{enumerate}
\item Neutrino-neutrino interaction.
\begin{eqnarray}
\frac{d\Gamma}{dx}&=&\frac{1}{2^7\pi^5}G^2F^2f_K^2m_K(1+r^2-2x)(x^2-r^2)^{1/2}\times 
\nonumber \\
& &\times[(1-2x)x+r^2],
\label{eq:nunu}
\end{eqnarray}
where $r=m_\mu/m_K$, $x=E_\mu/m_K$, $E_\mu$ is the total muon energy, $m_\mu$ and $m_K$ 
are muon and kaon masses respectively, $G$ is the Fermi constant, $F$ is the hypothetical neutrino-
neutrino interaction constant, and $f_K$ is kaon decay constant.  
\item Six fermion interaction.
\begin{eqnarray}
\frac{d\Gamma}{dE_\mu}&=&f_K^2F_S^22^9R\left(1+\frac{m_\mu}{E_\mu}\right), \nonumber \\
R&=&\frac{m_Kp_\mu A^2E_\mu}{3\pi^52^{13}}(2p_\mu^3+4p_\mu^2A+2.5p_\mu A^2+0.5A^3), 
\nonumber \\
A&=&m_K-E_\mu-p_\mu,
\label{eq:sixf}
\end{eqnarray}
where $f_K$ is kaon decay constant, $E_\mu$ is the total muon energy, $p_\mu$ is muon 
momentum, $m_K$ is kaon mass, $F_S$ is the common form factor which can be related to the 
usual four-fermion interaction constant $G$ by the expression:
\begin{equation}
F_S=\frac{G}{\sqrt{2}}\frac{1}{\lambda^3}.
\label{eq:fs}
\end{equation}
The constant $\lambda$ is an unknown parameter with the dimension of mass. 
\end{enumerate}
According to~\cite{StandardModel} and Eq.~(\ref{eq:nunu}, \ref{eq:sixf}) the differential decay rates 
for $K^+\to\mu^+\nu\bar\nu\nu$ decay are shown in Fig.~\ref{fig:models}. To compare the shape of 
the theoretical curves all spectra are normalized to 1.
 
The only direct search for the $K^+\to\mu^+\nu\bar\nu\nu$ decay was done by C.~Y.~Pang \textit{et 
al.}~\cite{pang}. The muon kinetic energy region $60<T_\mu<100$~MeV was 
examined~(corresponding to muon momentum region $127.6<p_\mu<176.4$~MeV/c).  No clear 
evidence of the signal decay was observed and a 90\%~C.L. upper limit was presented:

\begin{equation}
\frac{\Gamma(K^+\to\mu^+\nu\bar\nu\nu;60<T_\mu<100\textnormal{ MeV})}{\Gamma(K^+\to 
\textnormal{all})}<2.2\times 10^{-6}
\label{eq:pang_indep}
\end{equation}

The 90\%~C.L. upper limits on total decay rate were obtained assuming the muon spectrum from 
neutrino-neutrino interaction and six-fermion interaction models:

\begin{enumerate}

\item Neutrino-neutrino interaction:
\begin{equation}
\frac{\Gamma(K^+\to\mu^+\nu\bar\nu\nu)}{\Gamma(K^+\to \textnormal{all})}<6.0\times 10^{-6}; 
\label{eq:pang_nunu}
\end{equation}

and

\item Six-fermion interaction:
\begin{equation}
\frac{\Gamma(K^+\to\mu^+\nu\bar\nu\nu)}{\Gamma(K^+\to \textnormal{all})}<6.7\times 10^{-6}.
\label{eq:pang_sixf}
\end{equation}
\end{enumerate}

The Eq.~(\ref{eq:pang_nunu}) is the best current upper limit on the $K^+\to\mu^+\nu\bar\nu\nu$ 
decay rate. 

In this paper, we present the result of a search for the rare kaon decay $K^+\to\mu^+\nu\bar\nu\nu$ 
in the muon momentum region $130<p_\mu<175$~MeV/c using stopped kaon decay data from 
experiment E949 at Brookhaven National Laboratory~(BNL)~\cite{e949}. In this analysis we used 
data taken from March to June in 2002. The total exposure for this analysis is $1.70\times 
10^{12}$ stopped kaons~\footnote{This is slightly less than $1.71\times 10^{12}$ stopped kaons 
used for the E949 analysis~\cite{e949}.}.
This analysis is based on 
the search for heavy neutrinos, $\nu_H$, in the $K^+\to\mu^+\nu_H$ decays~\cite{e949nuh} since in 
both cases a single muon must be identified.
In the following the method of calculation of rates for any 
assumed $\mu^+$ spectrum is also described.

\section{Experiment \label{sec:e949}}

The E949 experiment was aimed at a measurement of the rare kaon decay $K^+\to\pi^+\nu\bar\nu$~
\cite{e949} and other processes. Therefore, the principal trigger selection criteria were designed to 
select pions and reject muons. However, secondary muons were present in the data set due to 
inefficiencies in the pion selection criteria applied.

\subsection{Detector \label{subsec:e949detector}}

The  E949 $K^+$ beam was produced by a high-intensity proton beam from the
Alternating Gradient Synchrotron at BNL.
Protons were accelerated to a momentum of 21.5~GeV/c and hit a  platinum
production target.
 
\begin{figure}[]
\centering
\includegraphics[width=\columnwidth]{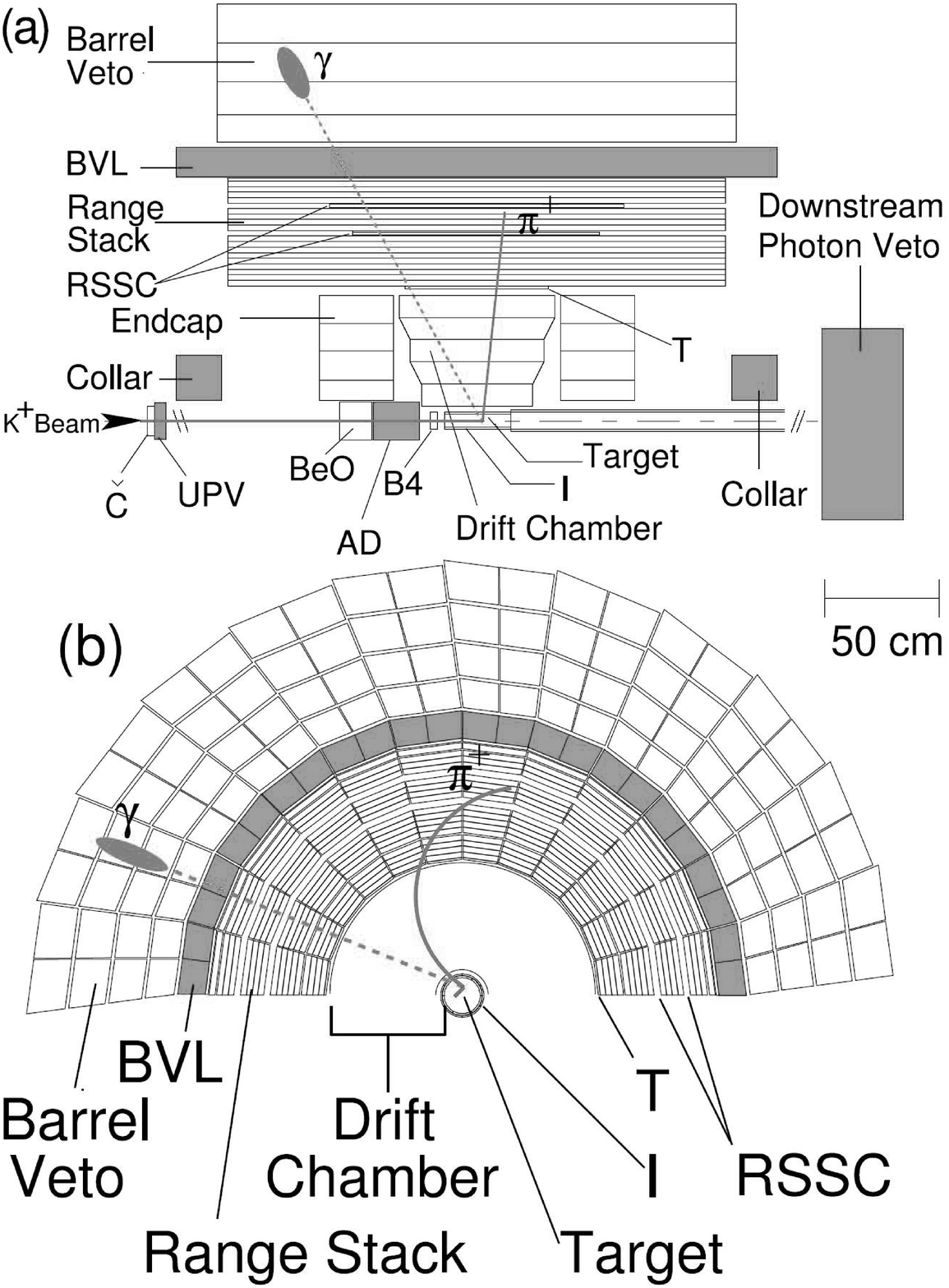}
\caption{Schematic side~(a) and end~(b) views of the upper half of the E949 detector. An incoming 
kaon is shown traversing the beam instrumentation, stopping in the target and decaying to $\pi^+
\pi^0$ . The outgoing charged pion and one photon from the $\pi^0\to\gamma\gamma$  decay are 
illustrated. Elements of the detector are described in the text.}
\label{fig:detector}
\end{figure}

The E949 detector is shown in Fig.~\ref{fig:detector}.
Incoming 710~MeV/c kaons with a $K^+/\pi^+$ ratio of 3/1 were identified by a \v{C}erenkov counter.
Downstream of the \v{C}erenkov counter was the upstream photon veto~(UPV). The UPV was used 
to veto beam particles coincident with the time of the $K^+$ decay. To monitor the beam profile and 
identify multiple incoming particle two proportional wire chambers~(BWPCs) were used. Downstream 
of the BWPCs kaons were slowed down by a passive BeO degrader and an active degrader~(AD).   
After passing through the degraders, a beam hodoscope~(B4) detected the entrance position of 
beam particles into the target and identified it as a kaon by measuring the energy deposit. The 
slowed down kaons came to rest in the 
center of the target.

The target was a 12~cm diameter cylinder, 3.1~m in length, which was made of 413 5~mm square 
Bicron BCF10 scintillating fibers. These fibers ran in the direction parallel to the beam and had 
0.09~mm thick inactive cladding. Smaller fibers, known as edge fibers, were used to fill the gaps 
near the outer edge of the target. Each of the 5~mm fibers was connected to a Hamamatsu 
R1635-02 photomultiplier tube~(PMT). 
The edge fibers were multiplexed into groups of 12 and each group was read out by a single PMT. 
The PMT signals were sent to analog-to-digital converters~(ADCs), time-to-digital 
converters~(TDCs) and charge-coupled devices~(CCDs). 

Typical energy deposits in an individual target fiber were on the order of tens of MeV for kaons 
traveling along the fiber, and only $\sim 1$~MeV for pions~(muons) passing perpendicularly through 
a fiber. The fiducial region of the target was defined by two layers of six plastic scintillator counters 
surrounding the target. The inner scintillators~(IC) surrounded 
the target and tagged charged decay products before they entered the drift chamber. The IC was 6.4 
~mm thick with an inner radius of 6.0~cm and extended 24~cm from the upstream face of the target. 
The outer scintillators (VC) overlapped the downstream edge of the IC by 6 mm and 
detected particles that decayed downstream of the fiducial region of the target. The VC was 5~mm 
thick and 1.96~m long. To prevent gaps, the VC elements were rotated by $30^\circ$ with respect to 
the IC 
elements. Each IC and VC element was read out by an PMT whose signal was sent to 
ADC, TDC and a 500 MHz transient digitizer~(TD) based on a flash ADC.

The drift chamber, called the Ultra Thin Chamber~(UTC), was located just outside of the IC, with an 
inner radius of 7.85~cm, an outer radius of 43.31~cm and length of 51~cm.
The whole E949 spectrometer was in a 1~Tesla magnetic field. The primary functions of the UTC 
were the momentum measurement of charged particles and providing a match between the tracks in 
the target and the range stack explained in the next paragraph. 

The Range Stack~(RS) consisted of 19 layers of scintillator counters, azimuthally segmented into 24 
sectors, and position-sensitive straw 
chambers embedded after the 10th and 14th layers of scintillator, and was located just outside of the 
UTC with an inner radius of 45.08~cm and an outer 
radius of 84.67~cm. The RS provided energy and range measurements.
The innermost layer, T-counter, 
was 0.635~cm thick and 52~cm long. The T-counter defined the fiducial volume for the charged 
decay products and was thinner than the remaining RS layers to suppress the rate from photon 
conversions. Each of the T-counter scintillators was read out by fibers coupled to a PMT at both the 
upstream and downstream ends. Layers 2--18 were 1.905~cm thick and 1.82~m long and each 
scintillator was coupled through light guides to PMTs at each end. Layer 19 was used mainly to veto 
long range charged particles such as muons. This layer 
was 1.0~cm thick and had the same length and read out method as layers 2--18. 

The detection of any activity coincident with the charged track was important for suppressing the 
backgrounds for $K^+\to\mu^+\nu\bar\nu\nu$ decay. The E949 detector included a hermetic system 
of photon veto detectors.
Nearly every detector subsystem contributed to the photon veto. Detectors whose sole purpose was 
the detection of photon activity were the 
Barrel Veto~(BV), the Barrel Veto Liner (BVL), the upstream and downstream End Caps, the 
upstream and downstream Collars, the downstream Microcollar ($\mu$CO) and the Downstream 
Photon Veto. Detectors that were part of the photon veto system, but also served other purposes 
were the AD, target, IC, VC and RS. For a given event, the regions of the target, IC and RS 
traversed by the charged track were excluded from the photon veto. The BV and BVL with 
thicknesses of 14.3 and 2.29 radiation lengths~(r.l.) at normal incidence, respectively, provided 
photon 
detection over 2/3 of the $4\pi$ solid angle. The photon detection over the remaining 1/3 of the $4\pi
$ solid angle was provided by the other calorimeters in the region from $10^\circ$ to $45^\circ$ of the beam axis 
with a total thickness from 7 to 15 r.l.

\subsection{The $K^+\to\mu^+\nu\bar\nu\nu$ trigger \label{subsec:e949trigger}}
The experimental signature of the $K^+\to\mu^+\nu\bar\nu\nu$ decay is similar to the $K^+\to\mu^+
\nu_H$ decay~(one single charged track with no other detector activity). This motivates the use of 
the same trigger. To search for a heavy neutrino we used the main E949 trigger~\cite{e949nuh}. This 
trigger consisted of several requirements. First, a kaon
had to enter the target. 
To be sure that the kaon decayed at rest,  the secondary charged particle had to
leave the target at least 1.5~ns later than the kaon hit in the \v{C}erenkov
detector. The 3-body $K^+$ decays were suppressed by the RS layer requirements.
The charged particle had to reach at least the sixth layer of the RS~(Fid\&Range). Long
tracks~(e.g. from $K^+\to\mu^+\nu_\mu$ decays) were suppressed by the layer 19 veto
requirement.  There were also additional refined requirements of the charged
track range took into account the number of target fiber hits and the track's
downstream position in RS layers 3, 11, 12, 13 as well as the
deepest layer of penetration~(refined range). The charged track had to be within
the fiducial region of all traversed RS layers.

The main trigger included  the  online pion identification in the RS. It
required a signature of $\pi^+\to\mu^+\nu_\mu$ decay in the online-selected  stopping
counter. The $\mu^+$ from the $\pi^+\to\mu^+\nu_\mu$ decay-at-rest had the
kinetic energy of 4~MeV~(few~mm equivalent range in plastic scintillator) and
rarely exited the stopping counter. So, pion pulses in the stopping counter
recorded by transient digitizers had a double-pulse structure.
Despite the online pion identification requirement, some muons remained in the
final sample due to inefficiency. 

Events were rejected if any activity in the photon detectors with energy above a
threshold was detected. This condition removed events with photons. A similar
requirement in the RS was also applied. The 24 sectors of the RS were grouped into six; a group of 
4 
sectors was called a "hextant".
Only one hextant was allowed to have hits or two hextants if they were adjacent.
This rejected events with multiple tracks and events with photon activity in the
RS. 

A more detail description of the E949 experiment can be found in~\cite{e949,e949_tr}.

\section{Analysis \label{sec:an}}
\subsection{Event selection \label{subsec:evsel}}

\begin{figure}[]
\centering
\includegraphics[width=\columnwidth]{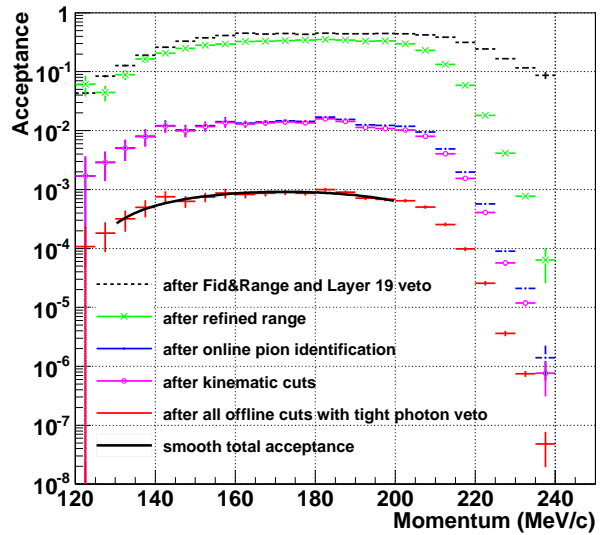}
\caption{Acceptance dependence on muon momentum. Black solid line shows the smooth total 
acceptance which is used for the further analysis.}
\label{fig:acc}
\end{figure}

In addition to the online selection criteria, offline selection criteria were also applied to suppress 
background processes and select single muon tracks. These criteria were exactly the same as for 
the heavy neutrinos search~\cite{e949nuh} since the experimental signature of the $K^+\to\mu^+\nu
\bar\nu\nu$ decay is similar to the $K^+\to\mu^+\nu_H$ decay. The acceptances of all selection 
criteria were measured using muon samples from the $K^+\to\mu^+\nu_\mu$ and $K^+\to\mu^+\nu_
\mu\gamma$ decays passed through monitor triggers. The description of acceptance measurements 
may be found in~\cite{e949nuh}. The total acceptance after all cuts is shown in Fig.~\ref{fig:acc}.
The acceptance drop off below 140~MeV/c is due to the requirement that the 
charged track must reach at least the sixth layer of the RS. 
The acceptance
drop off above 200~MeV/c is due to two requirements. First, the charged track
must not reach layer 19 of the range stack and second, the refined range
removes long tracks which are dominant at high momentum for events passing the
layer 19 requirement. The main acceptance loss~(factor $\sim 20$) comes from
the online pion identification requirement~(see~Figure~\ref{fig:acc}).

\subsection{The $K^+\to\mu^++X$ branching ratio}

\begin{figure*}[]
\centering
\includegraphics[width=15cm]{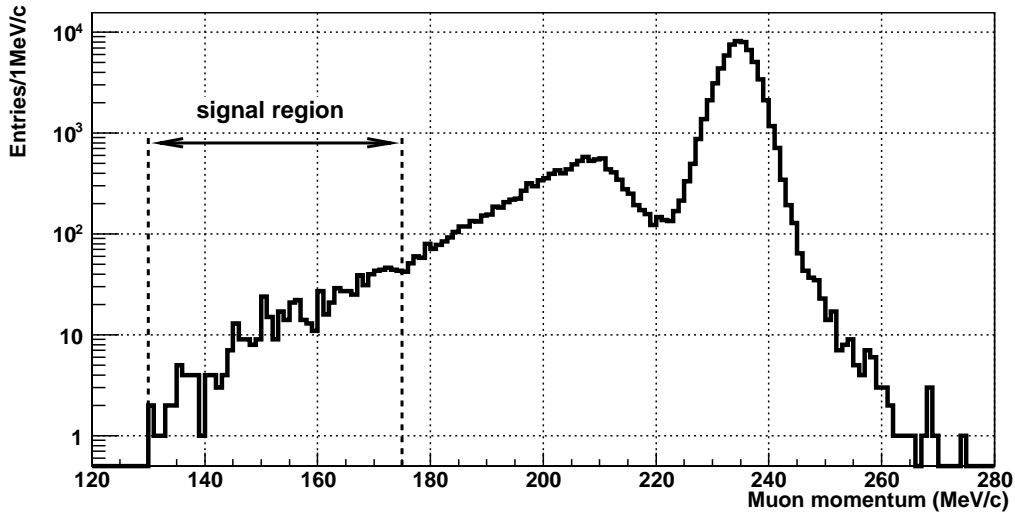}
\caption{Muon momentum spectrum for the full E949 data sample after all selection criteria were 
applied. The signal region is from 130 to 175~MeV/c.}
\label{fig:full}
\end{figure*}

\begin{figure*}[]
\centering
\includegraphics[width=15cm]{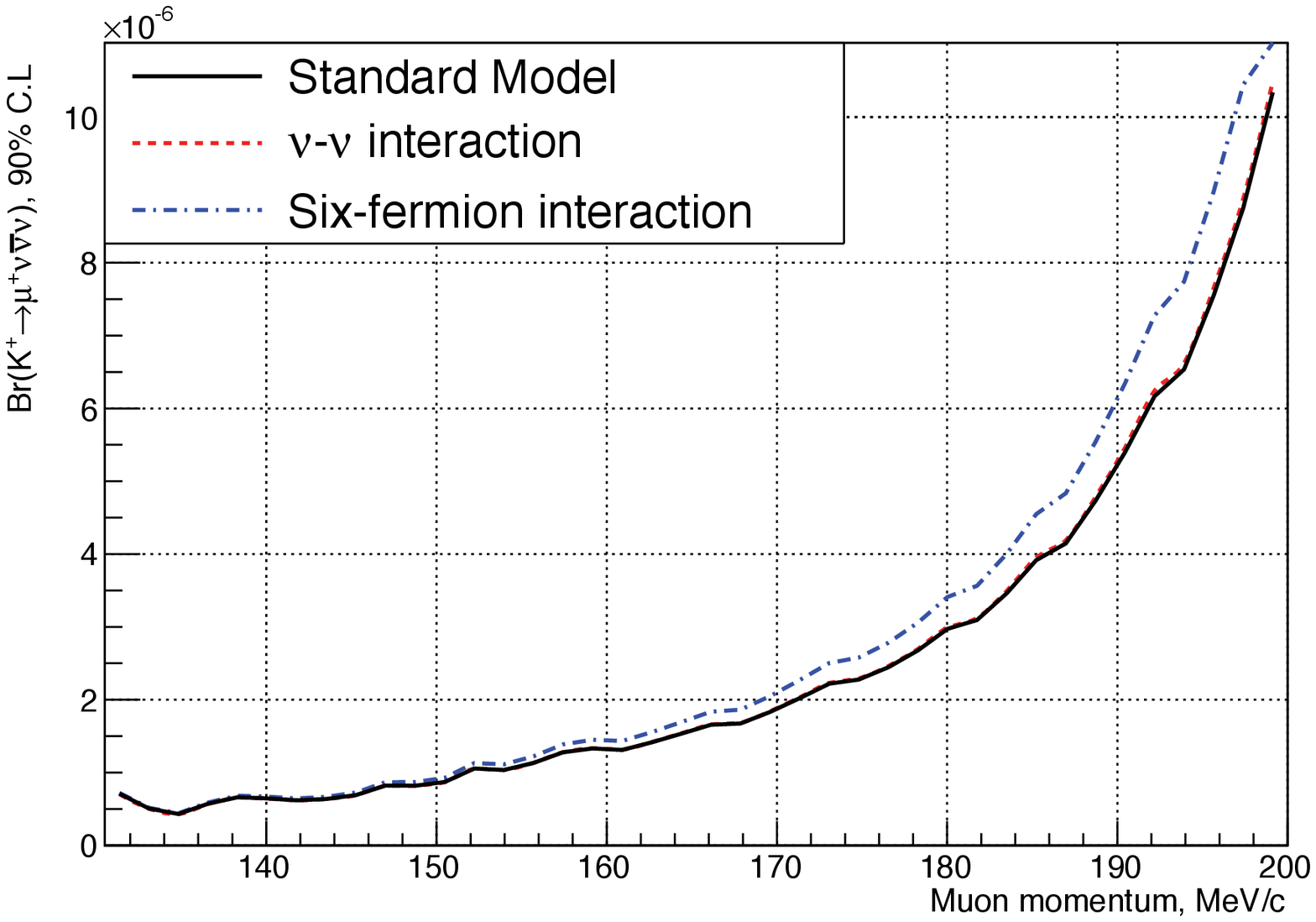}
\caption{Upper limits on the $K^+\to\mu^+\nu\bar\nu\nu$ branching ratio for the signal region 130-
P~MeV/c, where P is x-axis value.}
\label{fig:upperlimits_regions}
\end{figure*}

The muon momentum spectrum for the full E949 data sample after all selection criteria is shown in 
Fig.~\ref{fig:full}.
The data sample is dominated by the peak events at $p_\mu=236$~MeV/c~(muons from the $K^+\to
\mu^+\nu_\mu$ decay) and muons from the $K^+\to\mu^+\nu_\mu\gamma$ decay~\cite{e949nuh}.
Therefore, the main background source for the process $K^+\to\mu^++X$, where $X$ is set of 
neutral undetectable particles, is $K^+\to\mu^+\nu_\mu\gamma$ decay. Since we are not able to 
accurately predict the number of the background events after all selection criteria, we calculated the
standard one-sided 90\%~C.L. upper limits~(assuming gaussian distribution of measured values) on 
the signal process for different signal regions based 
on the 
observed events:
\begin{equation}
BR_P(K^+\to\mu^++X)<\frac{1}{N_K}\sum_{i=130}^{P}\frac{N_{i}}{Acc_i}
+1.28\sigma,
\label{eq:upperlimit}
\end{equation}
where $N_{i}$ is the number of observed events in the $i$th bin~(Fig.~\ref{fig:full}), $Acc_i$ is the 
total acceptance for the $i$th bin~(Fig.~\ref{fig:acc}), $N_K$ is the number of stopped kaons, 
$\sigma$ is the  total error which takes into account both uncertainties on $Acc_i$, $N_{i}$ and 
correlation between selected bins. The acceptance systematic uncertainty equals $\simeq 30\%$ of 
the acceptance value~\cite{e949nuh}. 
The lowest bound of the signal region, 130~MeV/c, was selected due to the acceptance drop 
off~(Fig.~\ref{fig:acc}). The upper bound of the signal region, $P$, was varied up to 200~MeV/
c~(limited by acceptance measurements).  To get total decay rate from Eq.~(\ref{eq:upperlimit}) we 
need to know the expected muon momentum spectrum. For any assumed spectrum the total decay 
rate on  $K^+\to\mu^+\nu\bar\nu\nu$ decay can be calculated using the following expression:
\begin{eqnarray}
\frac{\Gamma(K^+\to\mu^+\nu\bar\nu\nu)}{\Gamma(K^+\to \textnormal{all})}=BR_P(K^+\to\mu^++X)
\times \nonumber \\
\times\frac{\int_0^{p_{\mu}^{max}} (d\Gamma/dp_\mu)dp_\mu}{\int_{130}^{P} (d\Gamma/dp_\mu)dp_
\mu},
\label{eq:kmu3nutotal0}
\end{eqnarray}
where $BR_P(K^+\to\mu^++X)$ is defined by Eq.~(\ref{eq:upperlimit}), $p_{\mu}^{max}$ is the 
maximum muon momentum defined by kinematics~($p_\mu^{max}=(m_K^2-m_\mu^2)/2m_K$). 
Using the SM spectrum~\cite{StandardModel} and Eq.~(\ref{eq:nunu}, \ref{eq:sixf}, 
\ref{eq:kmu3nutotal0}), the total acceptance distribution from Fig.~\ref{fig:acc} and $N_K=1.70\times 
10^{12}$ we derived upper limits on the $K^+\to\mu^+\nu\bar\nu\nu$ branching ratio for different 
signal regions~(Fig.~\ref{fig:upperlimits_regions}).

\section{Results \label{sec:res}}

The number of observed events is drastically increased with muon momentum as shown in Fig.~
\ref{fig:full}. So, the upper bound of the signal region, 175~MeV/c, was selected to correspond to the 
previous experimental search for the $K^+\to\mu^+\nu\bar\nu\nu$ decay~\cite{pang}.  

Using Eq.~(\ref{eq:upperlimit}) with $P=175$~MeV/c we get the following upper limit on the partial 
$K^+\to\mu^++X$ branching ratio:
\begin{equation}
\begin{array}{ll}
BR(K^+\to\mu^++X,130<p_\mu<175\textnormal{ MeV/c})<\\
<7.5\times 10^{-7}
\end{array}
\label{eq:upperlimitindep}
\end{equation}

To get total decay rate from Eq.~(\ref{eq:upperlimitindep}) we used Eq.~(\ref{eq:kmu3nutotal0}):
\begin{equation}
\frac{\Gamma(K^+\to\mu^+\nu\bar\nu\nu)}{\Gamma(K^+\to \textnormal{all})}=7.5\times 10^{-7}\times 
\frac{\int_0^{p_{\mu}^{max}} (d\Gamma/dp_\mu)dp_\mu}{\int_{130}^{175} (d\Gamma/dp_\mu)dp_
\mu},
\label{eq:kmu3nutotal}
\end{equation} 
Using the SM spectrum~\cite{StandardModel} and Eq.~(\ref{eq:nunu}, \ref{eq:sixf}, 
\ref{eq:kmu3nutotal}) we get the following 90\%~C.L. upper limits on the total decay rate for $K^+\to
\mu^+\nu\bar\nu\nu$ decay:
\begin{enumerate}
\item Standard Model.
\begin{equation}
\frac{\Gamma(K^+\to\mu^+\nu\bar\nu\nu)}{\Gamma(K^+\to \textnormal{all})}<2.4\times 10^{-6}
\label{eq:e949_sm}
\end{equation}
\item Neutrino-neutrino interaction.
\begin{equation}
\frac{\Gamma(K^+\to\mu^+\nu\bar\nu\nu)}{\Gamma(K^+\to \textnormal{all})}<2.4\times 10^{-6}
\label{eq:e949_nunu}
\end{equation}

\item Six-fermion interaction:
\begin{equation}
\frac{\Gamma(K^+\to\mu^+\nu\bar\nu\nu)}{\Gamma(K^+\to \textnormal{all})}<2.7\times 10^{-6}
\label{eq:e949_sixf}
\end{equation}
\end{enumerate}
As can be seen from comparison of Eq.~(\ref{eq:e949_sm}, \ref{eq:e949_nunu}, \ref{eq:e949_sixf}) 
with Eq.~(\ref{eq:pang_nunu}, \ref{eq:pang_sixf}) the results
obtained with E949 data allow improvement of the limits on the $K^+\to\mu^+\nu\bar\nu\nu$ decay 
by about a factor of 3.

\section{Conclusion \label{sec:concl}}

A search for the rare decay $K^+\to\mu^+\nu\bar\nu\nu$ was done in the muon momentum region 
$130<p_\mu<175$~MeV/c using E949 data. No evidence of this process was found and we set new 
90\%~C.L. upper limit on the decay rate. Two proposed muon momentum spectra, neutrino-neutrino 
interaction and six-fermion interaction, were considered and we improved the current limits on the 
total decay rate in the framework of these models and for the Standard Model. We also presented 
the method of calculation of total rates on the $K^+\to\mu^+\nu\bar\nu\nu$ process for any assumed 
muon momentum spectrum.

\begin{acknowledgments}
This research was supported in part by Grant \#14-12-00560 of the Russian
Science Foundation, the U.S. Department of Energy, the Ministry of Education,
Culture, Sports, Science and Technology of Japan through the Japan-U.S.
Cooperative Research Program in High Energy Physics and under 
Grant-in-Aids for
Scientific Research, the Natural Sciences and Engineering Research Council~(Grant no. 157985) 
and the National Research Council of Canada, National Natural Science Foundation of
China, and the Tsinghua University Initiative Scientific Research Program.
\end{acknowledgments}

\bibliography{e949_kmu3nu-second-round}

\end{document}